\documentclass[twocolumn,superscriptaddress,letter]{revtex4}%
\usepackage{amssymb}
\usepackage{color}
\usepackage{graphicx}
\usepackage{dcolumn}
\usepackage{bm}
\usepackage[header,title,page,titletoc]{appendix}
\usepackage{amsmath}
\usepackage{amsfonts}%
\providecommand{\U}[1]{\protect \rule{.1in}{.1in}}
\begin{document}
\title{Strongly fluctuating fermionic superfluid in attractive $\mathrm{\pi}$-flux
Hubbard model}
\author{Ya-Jie Wu}
\affiliation{Department of Physics, Beijing Normal University, Beijing 100875, China}
\author{Jiang Zhou}
\affiliation{Department of Physics, Beijing Normal University, Beijing 100875, China}
\author{Su-Peng Kou}
\thanks{Corresponding author}
\email{spkou@bnu.edu.cn}
\affiliation{Department of Physics, Beijing Normal University, Beijing 100875, China}

\begin{abstract}
Ultracold atoms in optical lattice provides a platform to realize the
superfluid (SF) state, a quantum order with paired charge-neutral fermions. In
this paper, we studied SF state in the two-dimensional attractive Hubbard
model with $\pi$-flux on each plaquette. The SF state in the $\pi$-flux
lattice model suffers very strong quantum fluctuations and the ground state
becomes a possible \emph{quantum phase liquid} state. In this phase, there
exists the Cooper pairing together with a finite energy gap for the atoms, but
no long range SF phase coherence exists at zero temperature. In addition, we
discussed the properties of the SF vortices.

PACS number(s): 03.75.Ss, 67.85.Lm, 37.10.Jk

\end{abstract}
\maketitle

\section{Introduction and motivation}

Using ultracold atoms that form Bose-Einstein Condensates (BEC) or Fermi
degenerate gases to do precise measurements and\ simulations of quantum
many-body systems, is quite impressive and has become a rapidly-developing
field\cite{Bloch's review,review}. Since ultracold atoms may be trapped in
optical lattices, together with tunable interaction via Feshbach resonance
technique\cite{fesh,fesh1}, there is a new playground to manipulate a quantum
many-body system with unprecedented accuracy. In particular, an extreme
physical limit can be reached that is beyond the condition in the condensed
matter physics. Recently, using Raman-assisted tunneling in an optical lattice
of cold atoms, a large tunable (staggered) magnetic field was realized in
experiments\cite{ai,ai1,fl}. The synthetic gauge field enlarges the
versatility of the use of ultracold atoms very much, allowing exploring new
types of quantum states. In Ref.\cite{zhai}, Zhai \textit{et al} studied the
superfluid (SF) on a square lattice with a uniform magnetic flux $2\pi p/q$
($p$ and $q$ are co-prime integer numbers) on each plaquette, and found that
with considering the magnetic translation symmetry of the mean field ansatz,
the Cooper pairs may have finite momenta.

For a special artificial staggered gauge field in a square lattice with time
reversal symmetry and translation symmetry, there is a $\pi$-flux on each
plaquette. See the illustration in Fig.1. Namely, due to the nontrivial
Aharonov--Bohm phases, the atoms will obtain an extra minus sign after moving
around each plaquette. When one considers the attractive interaction between
atoms in a $\pi$-flux lattice model, the ground state can be an SF. An
interesting issue arises "\emph{whether the SF in }$\pi$\emph{-flux lattice
model has exotic quantum properties beyond the conventional SF in a square
lattice without }$\pi$\emph{-flux}?"

In this paper we will study this issue. We find that the SF in the $\pi$-flux
lattice model suffers quite strong quantum fluctuations. The strongly
fluctuating SF state possesses exotic quantum orders and cannot be described
by Landau's symmetry breaking theory. We call it \emph{quantum phase liquid}
(QPL) state. In QPL state, the\textit{ }fermions are paired as
\begin{equation}
\left \langle \left \vert \hat{c}_{i\uparrow}^{\dag}\hat{c}_{i\downarrow}^{\dag
}\right \vert \right \rangle =\left \langle \left \vert \Delta_{0}e^{i\phi_{i}%
}\right \vert \right \rangle =\left \vert \Delta_{0}\right \vert \neq0.
\end{equation}
And the single quasi-particle's excitation has a finite energy gap. However,
due to strong quantum fluctuations, the system has no coherence, or the
correlation length is finite. Hence, although the amplitude of the SF order
parameter is finite, the SF\ order parameter is zero, i.e.,
\begin{equation}
\langle \hat{c}_{i\uparrow}^{\dag}\hat{c}_{i\downarrow}^{\dag}\rangle
=\left \langle \Delta_{0}e^{i\phi_{i}}\right \rangle =\left \vert \Delta
_{0}\right \vert \left \langle e^{i\phi_{i}}\right \rangle =0
\end{equation}
with considering the random phase ($\phi_{i}\neq$ constant) of the SF order in
the QPL phase. At finite temperature, QPL state becomes a fermionic system
with pseudo-energy-gap. However, the fermionic SF vortex in QPL has different
topological properties with pseudo-energy-gap at finite temperature, of which
the SF\ vortex always obeys bosonic statistics.

\begin{figure}[ptb]
\scalebox{0.33}{\includegraphics* [0.8in,1.5in][20.5in,6.8in]{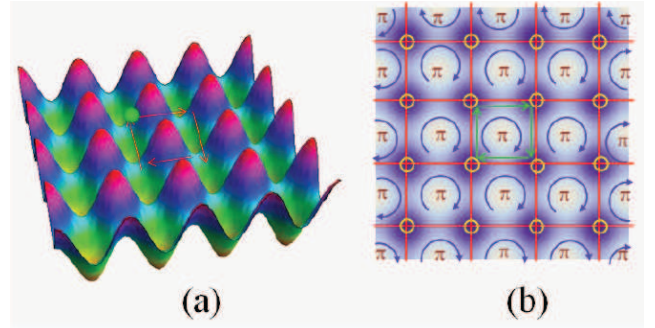}}\caption{(Color
online) (a) The illustration of potential of the two dimensional optical
lattice; (b) Hopping processes encircling a square plaquette acquire an
accumulated phase $\pi$. The background is contour plot of the optical
potential in (a).}%
\label{Fig1}%
\end{figure}

The reminder of this paper is organized as follows. In Sec. II, we write down
the $\pi$-flux attractive Hubbard model with a Zeeman field on square optical
lattice. Then, in Sec. III, we discuss the global symmetry of the system and
show the relationship between the attractive model and the repulsive one. In
Sec. IV, we obtain the phase diagram of the attractive model by the mean field
approach. In Sec. V, we investigate the phase fluctuations by using
random-phase-approximation\ (RPA) approach and obtain the dispersion of the
collective modes. In Sec. VI, we use the nonlinear $\sigma$ model (NL$\sigma
$M) method to derive the global phase diagram, and we find the QPL state in
the global phase diagram. In Sec. VII, we discuss the SF vortex and find that
the SF vortex in QPL state can be a fermionic excitation. Finally, we conclude
our discussions in Sec. VIII.

\section{The attractive Hubbard model in a $\pi$-flux square lattice}

Our starting point is the attractive Hubbard model on a $\pi$-flux square
lattice, of which the Hamiltonian is
\begin{align}
\hat{H}  &  =-\sum \limits_{\left \langle i,j\right \rangle ,\sigma}(t_{ij}%
\hat{c}_{i,\sigma}^{\dagger}\hat{c}_{j,\sigma}+h.c.)-U\sum \limits_{i}\hat
{n}_{i\uparrow}\hat{n}_{i\downarrow}\nonumber \\
&  -\mu \sum \limits_{i,\sigma}\hat{c}_{i\sigma}^{\dagger}\hat{c}_{i\sigma
}-h\sum \limits_{i,\sigma,\sigma^{\prime}}\hat{c}_{i\sigma}^{\dagger}%
\sigma_{\sigma \sigma^{\prime}}^{z}\hat{c}_{i\sigma^{\prime}}. \label{mod1}%
\end{align}
Here, $i=\left(  i_{x},i_{y}\right)  $ labels the lattice sites, and
$\left \langle i,j\right \rangle $ represents all nearest neighboring bonds. In
a $\pi$-flux lattice as shown in Fig.\ref{Fig1}, taking the Landau gauge as
example, the hopping parameters are $t_{i,i+\hat{e}_{x}}=t$ if $\left \langle
i,j\right \rangle $ is along $x$-direction and $t_{i,i+\hat{e}_{y}}%
=(-1)^{i_{x}}t$ if $\left \langle i,j\right \rangle $ is along $y$%
-direction\cite{zhai}. $\sigma,\sigma^{\prime}=\uparrow,\downarrow$ are
spin-indices, $U$ is the strength of the attractive interaction, $\mu$ is the
chemical potential, and $h$ is the strength of the Zeeman field. In the
following parts, we consider the case with $\mu=-U/2$ and take the lattice
constant $a$ equal to unity.

\section{Global symmetry}

Firstly, we discuss the global symmetry and the spontaneous symmetry breaking
of the original Hamiltonian in Eq.(\ref{mod1}). The Hamiltonian in
Eq.(\ref{mod1}) has an \textrm{SU(2)} particle-hole (pseudo-spin) symmetry
group, in which the \textrm{SU(2)} group elements act on the space of the
SF/CDW order parameters. To make the \textrm{SU(2)} pseudo-spin symmetry
clearer, we note that in terms of the canonical particle-hole
transformation\cite{mic}
\begin{equation}
\hat{c}_{i,\uparrow}\rightarrow \tilde{c}_{i,\uparrow}^{\dag},\hat
{c}_{i,\downarrow}\rightarrow(-1)^{i_{x}+i_{y}}\tilde{c}_{i,\downarrow}^{\dag
},
\end{equation}
the original model is mapped onto that of a repulsive $\pi$-flux model with
the effective chemical potential $h+U/2$. See Appendix A for the details. Then
we can define an \textrm{SU(2)} pseudo-spin symmetry of the Hamiltonian in
Eq.(\ref{mod1}), i.e.,
\begin{equation}
\hat{H}\rightarrow \hat{H}^{\prime}=\mathcal{U}\hat{H}\mathcal{U}^{-1}=\hat{H}%
\end{equation}
by doing a pseudo-spin rotation $\Psi \rightarrow \Psi^{^{\prime}}%
=\mathcal{U}\Psi,$ with $\Psi=(\tilde{c}_{i,\uparrow},\tilde{c}_{i,\downarrow
})^{T}$. The \textrm{SU(2)} pseudo-spin operators of the attractive $\pi
$-flux\textit{ }Hubbard model thus become
\begin{gather}
\hat{\eta}_{i}^{-}\leftrightarrow \left(  -1\right)  ^{i_{x}+i_{y}}\hat{\Delta
}_{i}=\left(  -1\right)  ^{i_{x}+i_{y}}\hat{c}_{i,\downarrow}\hat
{c}_{i,\uparrow},\text{ }\nonumber \\
\hat{\eta}_{i}^{+}\leftrightarrow \left(  -1\right)  ^{i_{x}+i_{y}}\hat{\Delta
}_{i}^{\dag}=\left(  -1\right)  ^{i_{x}+i_{y}}\hat{c}_{i,\uparrow}^{\dag}%
\hat{c}_{i,\downarrow}^{\dag},\\
\hat{\eta}_{i}^{z}\leftrightarrow(\hat{\rho}_{i}-1)/2,\nonumber
\end{gather}
where $\hat{\eta}_{i}^{\pm}=\hat{\eta}_{i}^{x}\pm i\hat{\eta}_{i}^{y},$ and
$\hat{\rho}_{i}$ is the particle density operator. One may check the
\textrm{SU(2)} algebraic relation between the \textrm{SU(2)} pseudo-spin
operators as
\begin{equation}
\lbrack \hat{\eta}_{i}^{\alpha},\hat{\eta}_{i}^{\beta}]=i\epsilon_{\alpha
\beta \gamma}\hat{\eta}_{i}^{\gamma}%
\end{equation}
with $\epsilon_{\alpha \beta \gamma}$ being antisymmetric tensor. Due to the
\textrm{SU(2)} pseudo-spin rotation symmetry, the ground state of the
Hamiltonian in Eq.(\ref{mod1}) can also be a charge density wave (CDW) state
as $\left \langle \hat{\eta}^{z}\right \rangle \neq0$.

For a conventional SF order from spontaneous \textrm{U(1)} phase symmetry
breaking, there exists one Goldstone mode due to the quantum phase
fluctuation. In two dimensions, there is a Kosterlitz-Thouless (KT)
transition, below which the (quasi-) long range phase coherence establishes.
Now we have an SF/CDW order from spontaneous \textrm{SU(2)} pseudo-spin
rotation symmetry breaking. The quantum fluctuations around the mean field
ground state are much stronger. In the CDW order one has nonzero particle
density modulation at different sublattices $\left \langle \hat{\eta}%
^{z}\right \rangle \neq0$. From the commutation relation between the phase
$\phi_{i}$ of $\Delta_{i}$ and the particle density operator $\hat{\rho}_{i}$,
i.e., $[\phi_{i},\hat{\rho}_{i}]\neq0$, the nonzero particle density
modulation leads to an uncertainty for the SF phase coherence and even
destroys the long range SF phase coherence. Thus, the non-zero value of
$\Delta_{0}$ only means the existence of Cooper pairing. It does not
necessarily imply that the ground state is a long range SF order. As a result,
one needs to examine the stability of SF order against quantum fluctuations
based on a formulation by keeping \textrm{SU(2)} pseudo-spin rotation symmetry.

\section{Mean field calculation}

For the attractive $\pi$-flux Hubbard model described by Eq.(\ref{mod1}), with
increasing the interaction strength, the ground state turns into an SF order.
Due to the existence of $\pi$-flux on each plaquette, we need to consider all
Cooper channels including zero momentum paring $\mathbf{q}=\left(  0,0\right)
$ and non-zero momentum paring $\mathbf{q}=\left(  0,\pi \right)  $. From
Ref.\cite{zhai}, the order parameter is written as
\begin{equation}
\Delta_{i}=\sum_{n=0}^{1}\Delta_{i_{x}(\operatorname{mod}2)}^{n}e^{i\pi
li_{y}}.
\end{equation}
Here $\Delta_{i_{x}(\operatorname{mod}2)}^{n}$ is the SF order with paring
momentum $\mathbf{q}=\left(  0,\pi n\right)  $ on sublattice site
$i_{x}(\operatorname{mod}2)$. For instance, sublattice $A$ refers to
$i_{x}(\operatorname{mod}2)=1$ and $B$ refers to $i_{x}(\operatorname{mod}%
2)=0$. The explicit form of the order parameters is given by $\Delta_{i\in
A,o}=\Delta_{i\in B,o}=a-ib$, and $\Delta_{i\in A,e}=\Delta_{i\in B,e}=a+ib$,
where the subscripts $o$ and $e$ correspond to odd rows and even rows of the
$\pi$-flux lattice, respectively.\begin{figure}[ptbh]
\includegraphics[width = 9.0cm]{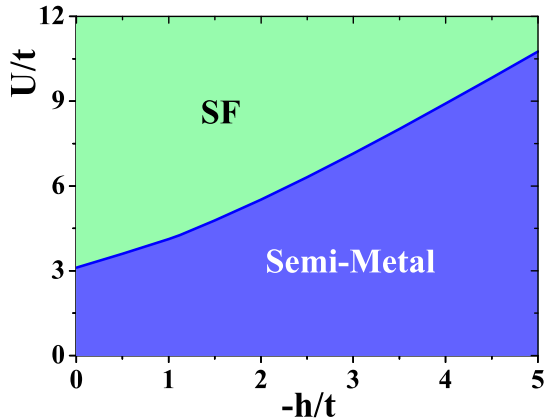}\caption{(Color online) Mean-field
phase diagram of the attractive $\pi$-flux Hubbard model without considering
phase fluctuations at zero temperature. There are two phases: semi-metal (blue
region), superfluid (green region). }%
\label{Fig2}%
\end{figure}

\begin{figure}[ptbh]
\includegraphics[width = 9.0cm]{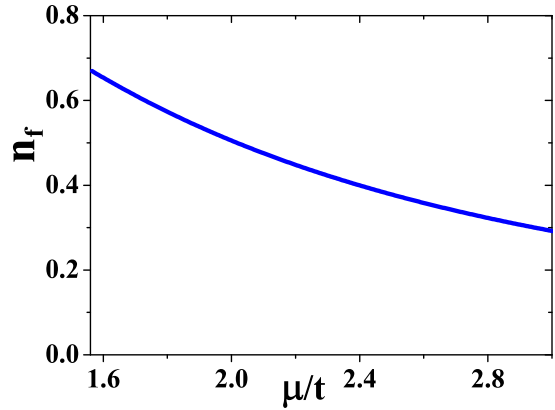}\caption{(Color online) The
illustration of particle-filling-number $n_{f}$ versus the chemical potential
$\mu$ for the case of Zeeman field $h=0$. }%
\label{Fig3}%
\end{figure}

By the numerical calculations, we find $b=0$ and the SF order parameter is
uniform, i.e., $\Delta_{i}=a=\Delta_{0}/2$. Then, by minimizing the free
energy $F$ with respect to $\Delta_{0}$, we obtain the gap equation
\begin{equation}
\frac{1}{U}=\frac{1}{4N_{s}}\sum_{\alpha,E_{\alpha,\mathbf{k}}>-h}\frac
{1}{E_{\alpha,\mathbf{k}}}\tanh \left(  \frac{\beta E_{\alpha,\mathbf{k}}}%
{2}\right)  ,
\end{equation}
where the energy spectra are
\begin{equation}
E_{\alpha,\mathbf{k}}=\sqrt{\xi_{\mathbf{\alpha},k}^{2}+\left(  m_{0}%
^{HS}\right)  ^{2}},
\end{equation}
with $\xi_{\alpha=\pm,\mathbf{k}}=2t\sqrt{[\cos^{2}(k_{x})+\cos^{2}(k_{y}%
)]}\pm \mu$ and $m_{0}^{HS}=U\Delta_{0}/2$, $\beta$ is given by $\beta
=1/\left(  k_{B}T\right)  $ with $k_{B}$ ($k_{B}\equiv1$) the Boltzmann
constant and $T$ the temperature, and $N_{s}$ is the number of the unit cells.
The summation is restricted in the reduced Brillouin zone where the relation
$E_{\mathbf{k}}>-h$ is satisfied. By solving the mean field equations, we plot
the phase diagram in Fig.\ref{Fig2} at zero temperature. The blue line in
Fig.\ref{Fig2} separates the gapped SF order and the semi-metal. In this
paper, the chemical potential is fixed as $\mu=-U/2$. We may identify the
particle-filling-number $n_{f}$ by $n_{f}=-\frac{1}{2N_{s}}(\frac{\partial
F}{\partial \mu})$. The particle-filling-number changes as variation of the
chemical potential $\mu$ (or interaction $U$). For example, for the case of
$h=0,$ we plot $n_{f}$ versus the chemical potential $\mu$ in Fig.\ref{Fig3}.
By this mean field theory, we can also get a finite SF transition temperature
$T^{\ast}$ corresponding to the temperature breaking the Cooper pair. In
general, this SF transition temperature is high. For example, for the case of
$h=-0.1t,$ $U=4t,$ the SF transition temperature is $T^{\ast}=0.75t$.

\section{Phase fluctuations by random-phase-approximation approach}

To study the quantum fluctuations of the SF/CDW order, firstly we use the
random-phase-approximation\ (RPA) approach to derive the dispersion of the
collective modes\cite{Iskin}.

In the imaginary-time functional integration formalism, we set $\beta=1/T$,
$\hbar=k_{B}=1$. The partition function is then written as%
\[
Z=\int D\left[  c^{\dagger},c\right]  e^{-S},
\]
where the effective action $S$ is%
\begin{equation}
S=\int_{0}^{\beta}d\tau \left[  \sum_{n,\mathbf{k,}\sigma}c_{n,\mathbf{k}%
,\sigma}^{\dagger}\left(  \tau \right)  \partial_{\tau}c_{n,\mathbf{k},\sigma
}\left(  \tau \right)  +H\left(  \tau \right)  \right]  ,
\end{equation}
where the term $H\left(  \tau \right)  $ can be readily obtained by replacing
the femionic operator $\hat{c}_{n,\mathbf{k,}\sigma}$ by the Grassman number
$c_{n,\mathbf{k,}\sigma}$.

By the Hubbard-Stratanovich transformation, the interaction term becomes
$e^{-S_{att}}$, where $S_{att}$ is%
\begin{align}
S_{att}  &  =\int_{0}^{\beta}d\tau \sum_{n,\mathbf{q,k}}[\tilde{\Delta
}_{n,\mathbf{q}}\left(  \tau \right)  c_{n,\mathbf{k+}\frac{\mathbf{q}}%
{2},\uparrow}^{\dagger}\left(  \tau \right)  c_{n,-\mathbf{k+}\frac{\mathbf{q}%
}{2},\downarrow}^{\dagger}\left(  \tau \right) \nonumber \\
&  +\tilde{\Delta}_{n,\mathbf{q}}^{\ast}\left(  \tau \right)  c_{n,-\mathbf{k}%
+\frac{\mathbf{q}}{2},\downarrow}\left(  \tau \right)  c_{n,\mathbf{k}%
+\frac{\mathbf{q}}{2},\uparrow}\left(  \tau \right) \nonumber \\
&  +\frac{N_{s}}{U}\sum_{n,\mathbf{q}}\tilde{\Delta}_{n,\mathbf{q}}^{\ast
}\left(  \tau \right)  \tilde{\Delta}_{n,\mathbf{q}}\left(  \tau \right)  ].
\end{align}
Performing an integration over the fermionic field, we have
\begin{equation}
S=\sum_{n,k}\left(  -i\omega_{l}-\mu \right)  +\frac{N_{s}}{U}\sum_{n,q}%
\tilde{\Delta}_{n,q}^{\ast}\tilde{\Delta}_{n,q}-tr\left[  \ln \left(
-G^{-1}\right)  \right]  ,
\end{equation}
where $i\omega_{l}=i\omega_{l}^{\prime}+h$ with $\omega_{l}^{\prime}=\left(
2l+1\right)  \pi/\beta,$ and the inverse of the Green function is given by
\begin{equation}
G^{-1}=i\omega_{l}\mathbb{I+}\left(
\begin{array}
[c]{cccc}%
-y_{\mathbf{k}}+\mu & x_{\mathbf{k}} & -\tilde{\Delta}_{A,q} & 0\\
x_{\mathbf{k}} & y_{\mathbf{k}}+\mu & 0 & -\tilde{\Delta}_{B,q}\\
-\tilde{\Delta}_{A,q}^{\ast} & 0 & y_{\mathbf{k}}-\mu & -x_{\mathbf{k}}\\
0 & -\tilde{\Delta}_{B,q}^{\ast} & -x_{\mathbf{k}} & -y_{\mathbf{k}}-\mu
\end{array}
\right)  , \label{G1}%
\end{equation}
where $\mathbb{I}$ is $4\times4$ identity matrix, the subscripts
$n=A$\textrm{, }$B$ describe two sublattices, and $x_{\mathbf{k}}=2t\cos
k_{x}$, $y_{\mathbf{k}}=2t\cos k_{y}$ (The lattice constant is defined as
$a\equiv1$)\textbf{.}

We split the $\tilde{\Delta}_{n,q}$ into a time-independent (stationary) part
(that is the mean field value) $\tilde{\Delta}_{n,0}$ and time-dependent part
(that represents the phase fluatuations) $\Lambda_{n}\left(  q\right)  $ as%
\begin{equation}
\tilde{\Delta}_{n,q}=\tilde{\Delta}_{n,0}\delta_{q,0}+\Lambda_{n}\left(
q\right)  ,
\end{equation}
where $q=\left(  \mathbf{q,}i\upsilon_{l}\right)  $ with $\upsilon_{l}%
=2l\pi/\beta.$ Then the Green function shown in Eq.(\ref{G1}) can be described
by
\begin{equation}
G^{-1}=G_{0}^{-1}+G_{1}^{-1},
\end{equation}
where $G_{0}^{-1}\left(  k,k\right)  $ describing the saddle point inverse
Nambu matrix takes the form as
\begin{equation}
G_{0}^{-1}=G^{-1}(\tilde{\Delta}_{A,q}^{\ast}\rightarrow \tilde{\Delta}%
_{A,0},\tilde{\Delta}_{B,q}\rightarrow \tilde{\Delta}_{B,0}).
\end{equation}
In the mean-field approach, the order parameters are $\tilde{\Delta}%
_{A,0}=\tilde{\Delta}_{B,0}=-m_{0}^{HS}$.

Expanding the term $tr\left[  \ln \left(  -G^{-1}\right)  \right]  $ by using
the Taylor formula up to the second order term, the effective action $S$ then
becomes $S=S_{0}+S_{2}$, where the zeroth order effective action is%
\begin{equation}
S_{0}=\sum_{n,k}\left(  -i\omega_{l}-\mu \right)  +\frac{N_{s}}{U}\sum
_{n,q}\tilde{\Delta}_{n,q}^{\ast}\tilde{\Delta}_{n,q}-tr\ln \left(  -G_{0}%
^{-1}\right)  ,
\end{equation}
and the second order effective action is
\begin{equation}
S_{2}=\frac{1}{2}tr\left[  G_{0}\left(  k+q\right)  \left(  G_{1}^{-1}\right)
_{k+q,k}G_{0}\left(  k\right)  \left(  G_{1}^{-1}\right)  _{k,k+q}\right]  .
\end{equation}
Then an effective action $\bar{S}_{flu}$ of quantum fluctuations $\Lambda
_{n}\left(  q\right)  $ becomes
\begin{equation}
\bar{S}_{flu}=\frac{\beta}{2}\sum_{q}\Lambda^{\dagger}\left(  q\right)
\Pi \left(  q\right)  \Lambda \left(  q\right)  =S_{2}+g\mathbb{I},
\end{equation}
with $\Lambda^{\dagger}\left(  q\right)  =\left(
\begin{array}
[c]{cccc}%
\Lambda_{A,q}^{\ast} & \Lambda_{A,-q} & \Lambda_{B,-q} & \Lambda_{B,q}^{\ast}%
\end{array}
\right)  .$

\begin{figure}[ptb]
\scalebox{0.43}{\includegraphics* [1.4in,2.1in][25.5in,6.5in]{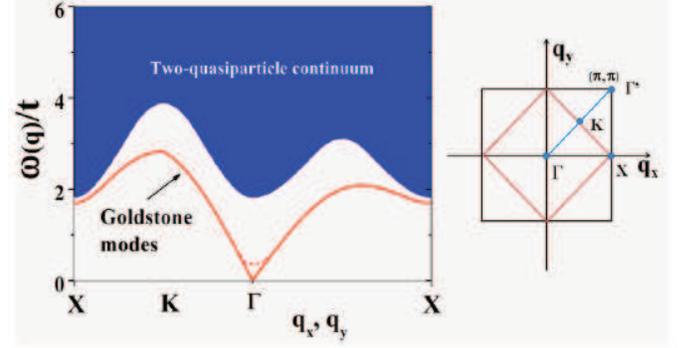}}\caption{(Color
online) The illustration of collective modes of SF/SDW order for the case of
$U/t=4.0$, $h=0.0$. The red line corresponds to the Goldstone modes. The blue
region denotes two-quasiparticle continuum. The Leggett modes merge into the
two-quasiparticle continuum. The red dotted line denotes a dispersion of the
Goldstone modes with a small energy gap obtained from the results of
\textrm{O(3)} nonlinear $\sigma$ model by renormalization-group approach.}%
\label{Fig4}%
\end{figure}

Next, we make use of the Matsubara summation formula to derive the explicit
form of $\Pi \left(  q\right)  $. For each element of the function $\Pi
_{ij}\left(  q\right)  ,$ we may write it as the summation of $\Pi
_{ij}^{qp-qp}\left(  q\right)  $ and $\Pi_{ij}^{qp-qh}\left(  q\right)  $,
where $\Pi^{qp-qp}\left(  q\right)  $ ($\Pi^{qp-qh}\left(  q\right)  $) is the
contribution term from the scattering between the quasi-particles and the
quasi-particles (the quasi-holes). In order to obtain the pair correlation
function (amplitude correlation function and phase fluctuation function), we
may first express the fluctuations of the order parameter as
\begin{equation}
\Lambda_{n}\left(  q\right)  =\eta_{n}\left(  q\right)  e^{i\phi_{n}\left(
q\right)  }=\left[  \lambda_{n}\left(  q\right)  +i\theta_{n}\left(  q\right)
\right]  /\sqrt{2}.
\end{equation}
where $\eta_{n}\left(  q\right)  $, $\phi_{n}\left(  q\right)  $, $\lambda
_{n}\left(  q\right)  $, $\theta_{n}\left(  q\right)  $ are all real fields.
$\lambda_{n}\left(  q\right)  =\sqrt{2}\eta_{n}\left(  q\right)  \cos \left[
\phi_{n}\left(  q\right)  \right]  $ and $\theta_{n}\left(  q\right)
=\sqrt{2}\eta_{n}\left(  q\right)  \sin \left[  \phi_{n}\left(  q\right)
\right]  $ can be essentially considered as the amplitude fluctuations and the
phase fluctuations, respectively. Then the vector $\Lambda \left(  q\right)
=\left(  \Lambda_{A,q},\Lambda_{A,-q}^{\ast},\Lambda_{B,-q}^{\ast}%
,\Lambda_{B,q}\right)  ^{T}$ is
\begin{equation}
\Lambda \left(  q\right)  =\frac{1}{\sqrt{2}}\left(
\begin{array}
[c]{cccc}%
1 & i & 0 & 0\\
1 & -i & 0 & 0\\
0 & 0 & -i & 1\\
0 & 0 & i & 1
\end{array}
\right)  \chi \left(  q\right)  ,
\end{equation}
where $\chi \left(  q\right)  =\left(
\begin{array}
[c]{cccc}%
\lambda_{A}\left(  q\right)  & \theta_{A}\left(  q\right)  & \theta_{B}\left(
q\right)  & \lambda_{B}\left(  q\right)
\end{array}
\right)  ^{T}$.

Because the low energy excitations are the phase fluctuations, we only focus
on the phase fluctuations in the following part. Integrating out the amplitude
fluctuations, the effective action of the phase fluctuations is obtained as
\begin{equation}
\bar{S}_{flu}\left(  \theta \right)  =\frac{\beta}{2}\sum_{q}[(%
\begin{array}
[c]{cc}%
\theta_{A} & \theta_{B}%
\end{array}
)\Theta \left(  q\right)  (%
\begin{array}
[c]{c}%
\theta_{A}\\
\theta_{B}%
\end{array}
)].
\end{equation}
In the static limit $i\upsilon_{l}\rightarrow \upsilon+i0^{+}$, the collective
modes at zero temperature then can be derived numerically from
\begin{equation}
\det \left[  \Theta \left(  q\right)  \right]  =0.
\end{equation}

See the results in Fig.\ref{Fig4}. From Fig.\ref{Fig4}, we find that there
exists a gapless collective mode corresponding to the Goldstone mode. In
addition, due to the two-sublattice there exists a gapped collective mode
corresponding to the Leggett mode. For the weakly coupling case, the Leggett
mode lies in the two-quasiparticle continuum and is strongly damped. With
increasing the coupling strength $U$, the two-quasiparticle continuum of the
Bogliubov qusiparticles is above the range of the Goldstone mode and the
Leggett mode, i.e., the undamped Leggett mode emerges.

\section{\textrm{O(3)} Nonlinear $\sigma$ Model}

However, the RPA approach underestimates the quantum fluctuations of the
SF/CDW\ order in the long-wave-length limit. To derive the quantum
fluctuations in the long-wave-length limit, we focus on the low energy physics
of the system by using the renormalization-group (RG) approach. Because the
amplitude fluctuations always have a large energy gap, we may ignore it and
consider the SF/CDW order parameter as an \textrm{O(3)} rotor with fixed
length as $\Delta_{0}/2$. \begin{figure}[h]
\includegraphics[width = 9.0cm]{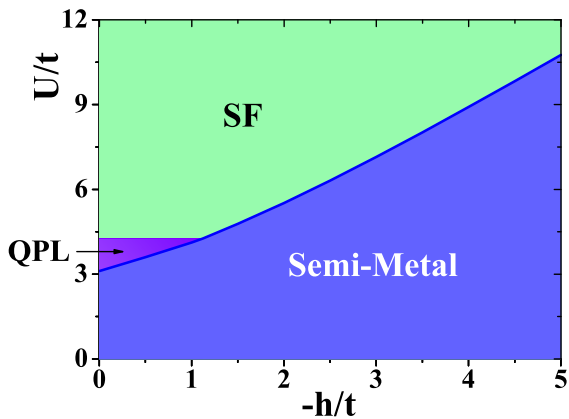}\caption{(Color online) Global phase
diagram of attractive $\pi$-flux Hubbard model with considering phase
fluctuations at zero temperature. There are three phases: semi-metal (blue
region), quantum phase liquid (purple region), superfluid (green region). }%
\label{Fig5}%
\end{figure}

Now, the effective Lagrangian with spontaneous \textrm{SU(2)} pseudo-spin
rotation symmetry breaking under the particle-hole transformation reads
\begin{align}
\mathcal{L}_{\mathrm{eff}}  &  =\sum_{i}\tilde{c}_{i}^{\dagger}\partial_{\tau
}\tilde{c}_{i}-\sum \limits_{\left \langle ij\right \rangle }(t_{i,j}\tilde
{c}_{i}^{\dagger}\tilde{c}_{j}+h.c.)\\
&  -\sum_{i}\left(  -1\right)  ^{i}m_{0}^{HS}\tilde{c}_{i}^{\dagger
}\mathbf{\Delta}_{i}\mathbf{\cdot \sigma}\tilde{c}_{i}-h\sum \limits_{i}%
\tilde{c}_{i}^{\dagger}\tilde{c}_{i}.\nonumber
\end{align}
The SC/CDW order of the attractive Hubbard model corresponds to an
antiferromagnetic order of the repulsive Hubbard model, and the quantum phase
fluctuations of the attractive Hubbard model correspond to quantum spin
fluctuations of the repulsive Hubbard model.

To describe the quantum fluctuations, we use the Haldane's mapping%
\begin{align}
\mathbf{\Delta}_{i}  &  =(\operatorname{Re}\Delta_{i},\operatorname{Im}%
\Delta_{i},(\rho_{i}-1)/2)\\
&  =\left(  -1\right)  ^{i}\mathbf{n}_{i}\Delta_{0}/2\sqrt{1-\mathbf{L}%
_{i}^{2}}+\mathbf{L}_{i},\nonumber
\end{align}
where $\mathbf{n}_{i}=(\frac{\operatorname{Re}\Delta_{i}}{\Delta_{0}/2}%
,\frac{\operatorname{Im}\Delta_{i}}{\Delta_{0}/2},\frac{\left(  -1\right)
^{i}(\rho_{i}-1)/2}{\Delta_{0}/2})$ is the \textrm{O(3)} rotor for the SF/CDW
order parameter corresponding to the long wavelength part of $\mathbf{\Delta
}_{i}$ with a restriction $\mathbf{n}_{i}^{2}=1$, and $\mathbf{L}_{i}$ is the
transverse canting field corresponding to the short wavelength part of
$\mathbf{\Delta}_{i}$ with a restriction $\mathbf{L}_{i}\cdot \mathbf{n}_{i}%
=0$. \ 

In the long-wave-length limit, after integrating out the fermions and the
transverse canting field, the collective modes of the SF/CDW order can be
described by the effective \textrm{O(3)} nonlinear $\sigma$-model (NL$\sigma
$M)\cite{Haldane,Dupuis}:
\begin{equation}
\mathcal{L}_{\mathrm{SF/CDW}}=\frac{1}{2gv}[\left(  \partial_{\tau}%
\mathbf{n}\right)  ^{2}+v^{2}\left(  \mathbf{\bigtriangledown n}\right)
^{2}].
\end{equation}
Here, the coupling constant $g$ and the collective mode's velocity $v$ are
defined as
\begin{equation}
g=\frac{v}{\rho_{\mathrm{phase}}},
\end{equation}
and
\begin{equation}
v^{2}=\rho_{\mathrm{phase}}[(\frac{1}{4N_{s}}\sum_{E_{\mathbf{k}}>-h}%
(m_{0}^{HS})^{2}/E_{\mathbf{k}}^{\frac{3}{2}})^{-1}-2U]
\end{equation}
where the phase stiffness of the SF order is \begin{widetext}
\begin{equation}
\rho_{\mathrm{phase}}=\frac{1}{2N_{s}}\sum_{E_{k}>-h}\frac{\{(m_{0}^{HS}%
)^{2}+3t^{2}+t^{2}\cos \left(  4k_{x}\right)  +\cos \left(  2k_{x}\right)
\left[  8t^{2}+4t^{2}\cos \left(  2k_{y}\right)  +(m_{0}^{HS})^{2}\right]
\}^{2}}{\left(  \xi_{k}^{2}+(m_{0}^{HS})^{2}\right)  ^{\frac{3}{2}}%
},\label{rou}%
\end{equation}
\end{widetext}and the energy spectrum $E_{\mathbf{k}}=\sqrt{\xi_{\mathbf{k}%
}^{2}+\left(  m_{0}^{HS}\right)  ^{2}}$ with $\xi_{\mathbf{k}}=2t\sqrt
{[\cos^{2}(k_{x})+\cos^{2}(k_{y})]}$. See Appendix B for the detailed calculations.

The properties of the effective \textrm{O(3)} NL$\sigma$M are determined by
the dimensionless coupling constant $\alpha=g\Lambda$\cite{cha,sech}, of which
the cutoff is defined as $\Lambda=\min \left(  1,\text{ }2m_{0}^{HS}/v\right)
$. Using the RG approach\cite{cha}, the RG scaling equation had been obtain
as
\begin{equation}
\frac{d\alpha}{dl}=-\alpha+\frac{\alpha^{2}}{4\pi}%
\end{equation}
where $e^{l}$ is the length rescaling factor. Particularly, there exists a
critical point $\alpha_{c}=4\pi$ ($g_{c}=\frac{4\pi}{\Lambda}$). The quantum
critical point separates the long range SF/CDW order and the short range one
(the quantum phase liquid). The global phase diagram with considering phase
fluctuations of the $\pi$-flux attractive Hubbard model is given in
Fig.\ref{Fig5}. From it, we can see that except for the semi-metal phase
($\Delta_{0}=0$) and the SF-CDW phase ($\Delta_{0}\neq0$, $\alpha<4\pi$),
there exists an additional phase ($\Delta_{0}\neq0$, $\alpha>4\pi$) - quantum
phase liquid (the purple region in Fig.\ref{Fig5}). For the case of $h=0$, the
quantum phase liquid lies between $\left(  U/t\right)  _{c_{1}}=3.12$ and
$\left(  U/t\right)  _{c_{2}}=4.26$. As the Zeeman field strength increases,
the QPL region shrinks. For the case of $-h=0.5t$, the region of the QPL is
$\left(  U/t\right)  _{c_{1}}=3.60<U/t<\left(  U/t\right)  _{c_{2}}=4.26$.
While for $-h>1.06t$, there doesn't exist the QPL at all.

\begin{figure}[ptbh]
\includegraphics[width =8.6cm]{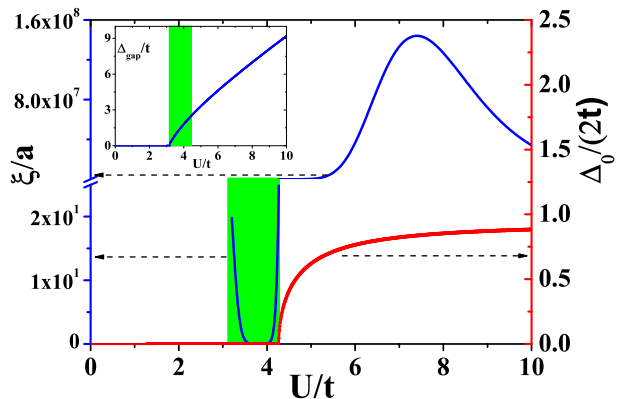}\caption{(Color online) The SF
correlation length $\xi$ (the blue line) at $k_{B}T=0.02t$ with $h=-0.1t$ and
the SF order parameter (the red line) at zero temperature with $h=-0.1t$. The
green region is the QPL with short range SF correlation and zero SF order
parameter. The inset is the SF energy gap of fermions which is finite in the
QPL at zero temperature.}%
\label{Fig6}%
\end{figure}

For the case of $\Delta_{0}\neq0$, $\alpha<4\pi$, at zero temperature, the
interaction between the collective modes is irrelevant ($g(l)\rightarrow0$ for
$l\rightarrow \infty$) and the ground state has long range SF/CDW order. As
shown in Fig.\ref{Fig6} (the red line), the order parameter is not zero
$\langle \hat{c}_{i\uparrow}^{\dag}\hat{c}_{i\downarrow}^{\dag}\rangle \neq0$.
However, due to the strong thermal fluctuations, the SF transition temperature
is zero. Fig.\ref{Fig6} (the blue line) also shows the SF correlation length
$\xi$ at $k_{B}T=0.02t$ \cite{exp}which is really an infinite value for
$T\rightarrow0$.

For the case of $\Delta_{0}\neq0$, $\alpha>4\pi$, the quantum fluctuations are
strong enough and the interaction between the collective modes becomes
relevant ($g(l)\rightarrow \infty$ for $l\rightarrow \infty$). Now the ground
state turns into QPL. There is no long range SF phase coherence and the SF
order parameter is zero, $\langle \hat{c}_{i\uparrow}^{\dag}\hat{c}%
_{i\downarrow}^{\dag}\rangle=0$. Thus in this region, the SF correlation
decays exponentially
\begin{equation}
\left \langle \hat{\Delta}^{\ast}(x,y)\hat{\Delta}(0)\right \rangle \sim
e^{i\mathbf{q}\cdot \mathbf{x}_{i}}e^{-r/\xi}%
\end{equation}
with $\mathbf{q}=(\pi,\pi)\mathbf{,}$ $r=\sqrt{x^{2}+y^{2}}$. Here $\xi$ is
the SF correlation length, $\xi=[4\pi v(\frac{1}{g_{c}}-\frac{1}{g})]^{-1}.$
Then the collective excitations have a mass gap as $4\pi v\left(  \frac
{1}{g_{c}}-\frac{1}{g}\right)  $ (see the red dotted line in Fig.\ref{Fig4},
of which a small energy gap of the Goldstone modes appears due to quantum
fluctuations). The SF correlation length $\xi$ is a finite value as
$T\rightarrow0$. The inset of Fig.\ref{Fig6} shows that the energy gap of the
paired fermions is always finite in the QPL region.

It is necessary to point out that the QPL corresponds to quantum spin liquid
as shown in Fig.\ref{Fig8}(b) in the repulsive $\pi$-flux Hubbard model on
square lattice in the intermediate coupling region\cite{kou1}. The prediction
of the quantum spin liquid state near Mott insulator (MI) transition of the
repulsive $\pi$-flux Hubbard model in Ref.\cite{kou1} has been confirmed by
the quantum Monte-Carlo (QMC) calculation\cite{pi}.

\section{Superfluid vortex}

\begin{figure}[ptb]
\scalebox{0.35}{\includegraphics* [0.8in,1.8in][23.8in,6.7in]{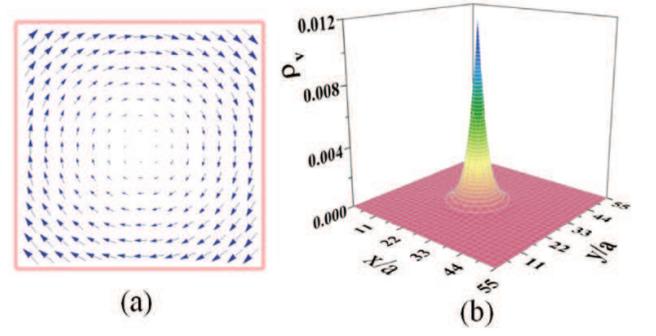}}\caption{(Color
online) (a) The illustration of an SF vortex. (b) The particle-density
$\rho_{v}$ of the fermionic zero modes around an SF vortex on 55-by-55
lattice. We take the case of $U=3.5t$, $\Delta_{0}=0.25$, $h=0.0$ as an
example.}%
\label{Fig7}%
\end{figure}

Finally we study the topological excitations in the QPL - the SF vortices. The
SF vortex solution is known as
\begin{equation}
\Delta_{i}=\Delta_{0}/2\exp \{ \pm i\mathrm{Im}[\ln(z_{i}-z_{0})]\phi_{i}\}
\label{ed}%
\end{equation}
where $z\equiv x+iy$ denotes the position of the SF vortex and the subscript
$i$ denotes\ the lattice site. See the illustration in Fig.\ref{Fig7}(a). In
long range SF/CDW order, SF vortex and SF anti-vortex have infinite energy and
are all confined. While in the short range SF/CDW order (QPL), the SF vortex
and SF anti-vortex have finite energy and are deconfined. Now the SF vortices
are true excitations. A question arises "\emph{is the vortex a boson or a
fermion}?" To answer this question we study the induced quantum number on the
vortex firstly.

\begin{figure}[h]
\scalebox{0.44}{\includegraphics* [1.80in,1.60in][9.70in,6.3in]{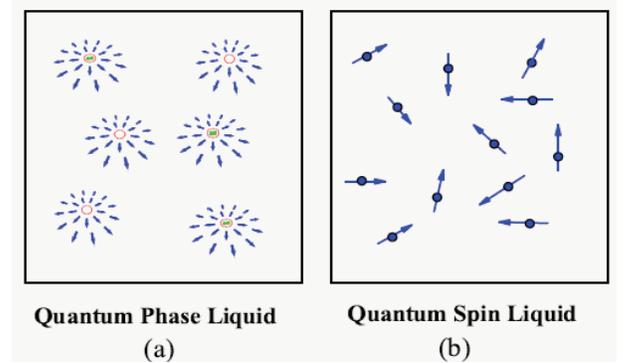}}\caption{(Color
online) The illustration of the quantum phase liquid (QPL) and quantum spin
liquid (QSL): (a) The elemental excitations of QPL are the two-component
fermionic SF vortices (or anti-vortices) with or without a Cooper pair inside
the vortex core. Two green spots denote a Cooper pair; (b) The elemental
excitations of the QSL are the spinons.}%
\label{Fig8}%
\end{figure}

From the numerical results, we find that there exist two fermionic zero modes
on each SF vortex. Fig.\ref{Fig7}(b) is the particle-density of the fermionic
zero modes around an SF vortex on $55$-by-$55$ lattice by the numerical
calculations. As shown in Ref.\cite{kou3}, the existence of the fermionic zero
mode leads to an induced pseudo-spin number inside the SF vortex core as
$\left \langle \hat{\eta}^{z}\right \rangle =\pm \frac{1}{2}$. According to the
mapping from the attractive model to the repulsive model, i.e., $\frac{1}%
{2}\left(  \rho-1\right)  $ $\leftrightarrow \hat{\eta}^{z}$, for $\left \langle
\hat{\eta}^{z}\right \rangle =\frac{1}{2},$ we have
\begin{equation}
\left \langle
%TCIMACRO{\dsum \limits_{i}}%
%BeginExpansion
{\displaystyle \sum \limits_{i}}
%EndExpansion
\left(  \hat{c}_{i\uparrow}^{\dagger}\hat{c}_{i\uparrow}+\hat{c}_{i\downarrow
}^{\dagger}\hat{c}_{i\downarrow}\right)  \right \rangle =2,
\end{equation}
which means a pair of fermions inside the SF vortex; for $\left \langle
\hat{\eta}^{z}\right \rangle =-\frac{1}{2},$ we have
\begin{equation}
\left \langle
%TCIMACRO{\dsum \limits_{i}}%
%BeginExpansion
{\displaystyle \sum \limits_{i}}
%EndExpansion
\left(  \hat{c}_{i\uparrow}^{\dagger}\hat{c}_{i\uparrow}+\hat{c}_{i\downarrow
}^{\dagger}\hat{c}_{i\downarrow}\right)  \right \rangle =0,
\end{equation}
which means such SF vortex is trivial. See the illustration in Fig.\ref{Fig8}(a).

In the quantum phase liquid, the superfluid (SF) vortex and the SF anti-vortex
have finite energy and are deconfined. Now the SF vortices are true
excitations. A question arises "\emph{is the vortex a boson or a fermion}?"
Let us answer this question.

Firstly, we calculate the fermion zero modes on the SF vortex by the continuum
formula of the effective model in Eq.(\ref{model2}) in the Appendix B. The SF
vortex solution is given in Eq.(\ref{ed}). The size of the vortex core is
$\Lambda^{-1}=\max(a,$ $(\Delta_{0})^{-1})$. After the particle-hole
transformation, the SF vortex of the attractive Hubbard model corresponds to
the half-skyrmion of the repulsive Hubbard model as
\begin{equation}
\mathbf{n}_{0}=(\frac{x-x_{0}}{|\mathbf{r}-\mathbf{r}_{0}|},\text{ }\pm
\frac{y-y_{0}}{|\mathbf{r}-\mathbf{r}_{0}|},\text{ }0).
\end{equation}
In the continuum limit, the effective Lagrangian describes the low energy
fermionic excitations at two nodes $\mathbf{k}_{1}=(\frac{\pi}{2},\frac{\pi
}{2}),$ $\mathbf{k}_{2}=(\frac{\pi}{2},-\frac{\pi}{2})$, and is written as
\begin{align}
\mathcal{L}_{\mathrm{eff}}  &  =i\bar{\Psi}_{1}\gamma_{\mu}\partial_{\mu}%
\Psi_{1}+i\bar{\Psi}_{2}\gamma_{\mu}\partial_{\mu}\Psi_{2}\nonumber \\
&  +m_{0}^{HS}(\bar{\Psi}_{1}\mathbf{n_{0}\cdot \sigma}\Psi_{1}-\bar{\Psi}%
_{2}\mathbf{n_{0}\cdot \sigma}\Psi_{2})
\end{align}
where $\Psi_{1}=(%
\begin{array}
[c]{cccc}%
\tilde{c}_{\uparrow1\mathrm{A}} & \tilde{c}_{\uparrow1\mathrm{B}} & \tilde
{c}_{\downarrow1\mathrm{A}} & \tilde{c}_{\downarrow1\mathrm{B}}%
\end{array}
)^{T}$ and $\Psi_{2}=(%
\begin{array}
[c]{cccc}%
\tilde{c}_{\uparrow2\mathrm{B}} & \tilde{c}_{\uparrow2\mathrm{A}} & \tilde
{c}_{\downarrow2\mathrm{B}} & \tilde{c}_{\downarrow2\mathrm{A}}%
\end{array}
)^{T}$ with $\mathrm{A}$ and $\mathrm{B}$ representing sublattices.
$\gamma_{\mu}$ is defined as $\gamma_{0}=\sigma_{0}\otimes \tau_{z},$
$\gamma_{1}=\sigma_{0}\otimes \tau_{y},$ $\gamma_{2}=\sigma_{0}\otimes \tau
_{x},$ $\sigma_{0}=\left(
\begin{array}
[c]{ll}%
1 & 0\\
0 & 1
\end{array}
\right)  $. $\tau^{x},$ $\tau^{y},$ $\tau^{z}$ are Pauli matrices. We set the
Fermi velocity to be unit, i.e., $v_{F}=1$. The solutions of zero modes are
given by%
\begin{equation}
\Psi_{1}^{0}(\mathbf{r})=\left(
\begin{array}
[c]{l}%
0\\
\exp(-\frac{\mid \mathbf{r}-\mathbf{r}_{0}\mid}{m_{0}^{HS}})\\
\exp(-\frac{\mid \mathbf{r}-\mathbf{r}_{0}\mid}{m_{0}^{HS}})\\
0
\end{array}
\right)  \text{, }\Psi_{2}^{0}(\mathbf{r})=\left(
\begin{array}
[c]{l}%
0\\
-\exp(-\frac{\mid \mathbf{r}-\mathbf{r}_{0}\mid}{m_{0}^{HS}})\\
\exp(-\frac{\mid \mathbf{r}-\mathbf{r}_{0}\mid}{m_{0}^{HS}})\\
0
\end{array}
\right)
\end{equation}
in Ref.\cite{kou3}. This result is consistent with that in Fig.\ref{Fig7}(b)
by numerical calculations.

Next, we calculate the induced quantum number on the SF vortex. For the
solutions of zero modes, there are four zero-energy soliton states around an
SF vortex which are denoted by
\begin{align}
&  \mid1_{+}\rangle \otimes \mid2_{+}\rangle,\text{ }\mid1_{-}\rangle \otimes
\mid2_{-}\rangle,\\
&  \mid1_{-}\rangle \otimes \mid2_{+}\rangle,\text{ }\mid1_{+}\rangle \otimes
\mid2_{-}\rangle.\nonumber
\end{align}
Here $\mid1_{-}\rangle$ and $\mid2_{-}\rangle$ are empty states of the zero
modes $\Psi_{1}^{0}(\mathbf{r})$ and $\Psi_{2}^{0}(\mathbf{r});$ $\mid
1_{+}\rangle$ and $\mid2_{+}\rangle$ are occupied states of them. At half
filling, the soliton states of an SF vortex $\mid \mathrm{sol}\rangle$ are
denoted by $\mid1_{-}\rangle \otimes \mid2_{+}\rangle$ and $\mid1_{+}%
\rangle \otimes \mid2_{-}\rangle$. In Ref.\cite{kou3}, the induced quantum
numbers of the solitons states including total induced fermion number $\hat
{N}_{F}=\sum \nolimits_{\alpha,i}\hat{c}_{i}^{\dagger}\sigma_{z}\hat{c}_{i}$
and the induced staggered spin number $\hat{\eta}_{(\pi,\pi)}^{z}=\frac{1}%
{2}\sum \nolimits_{i\in \mathrm{A}}\tilde{c}_{i}^{\dagger}\sigma_{z}\tilde
{c}_{i}-\frac{1}{2}\sum \nolimits_{i\in \mathrm{B}}\tilde{c}_{i}^{\dagger}%
\sigma_{z}\tilde{c}_{i}$ have been calculated. The total induced fermion
number on the solitons is zero due to the cancelation effect between two
nodes, i.e., $\hat{N}_{F}\mid \mathrm{sol}\rangle=0.$ However, there exists
an\ induced staggered pseudo-spin moment on the soliton states\cite{kou3},
\begin{equation}
\hat{\eta}_{(\pi,\pi)}^{z}\mid \mathrm{sol}\rangle=\pm \frac{1}{2}%
\mid \mathrm{sol}\rangle.
\end{equation}
From the fact of $\mathbf{n}_{i}=\mathbf{\bar{z}}_{i}\mathbf{\sigma z}_{i}$
where $\mathbf{z}$ is a bosonic spinon, $\mathbf{z}=\left(  z_{1}%
,z_{2}\right)  $ and $\mathbf{\bar{z}z=1}$, the induced staggered spin number
on a SF vortex means that there exists a trapped bosonic spinon inside the
vortex-core. On the other hand, the \textrm{O(3)} nonlinear $\sigma$ model is
equivalent to\textrm{\ CP(1)} model
\begin{align}
\mathcal{L}_{\mathbf{s}} &  =\frac{1}{2gv}[\left(  \partial_{\tau}%
\mathbf{n}\right)  ^{2}+v^{2}\left(  \mathbf{\bigtriangledown n}\right)
^{2}]\nonumber \\
&  =\frac{2}{gv}[|(\partial_{\tau}-ia_{\tau})\mathbf{z}|^{2}+v^{2}%
|(\vec{\nabla}-i\vec{a})\mathbf{z}|^{2}]
\end{align}
where $a_{\mu}\equiv-\frac{i}{2}(\mathbf{\bar{z}}\partial_{\mu}\mathbf{z}%
-\partial_{\mu}\mathbf{\bar{z}z})$\ is introduced as an auxiliary gauge field.
That means the bosonic spinon $\mathbf{z}$ carries a unit charge of the
auxiliary gauge field $a_{\mu}$. When the bosonic spinon $\mathbf{z}$ moves
around an SF vortex, its wave-function will obtain an extra minus sign. The SF
vortex is really a $\pi$-flux of the bosonic spinon $\mathbf{z}$. Hence, due
to the mutual semion statistics between the bosonic spinon $\mathbf{z}$ and
the SF vortex, a mobile SF vortex trapping a bosonic spinon $\mathbf{z}$
becomes a composite fermionic particle. We call such composite object (fermion
with $\pm \frac{1}{2}$ pseudo-spin degree freedom) a "\emph{fermionic SF
vortex}"\cite{kou3}.

Then, if there is no magnetic flux, i.e., $B=0,$ the density of the fermionic
SF vortices is zero at zero temperature. When we apply the extra magnetic
field, $B\neq0$ (or away from the $\pi$-flux case slightly), the density of
the fermionic SF vortices becomes finite\cite{comment}. In conventional SFs
with external magnetic field, below the Kosterlitz-Thouless (KT) transition
temperature, people can observe a vortex-lattice; While above the KT
transition temperature, the vortex-lattice will melt and the (bosonic)
vortices will move randomly. In the QPL, because each quantized magnetic flux
turns into a fermionic SF vortex, there is a dilute fermionic-SF-vortex gas
that forms a Fermi liquid for the weak magnetic field case. The fermionic
vortex density is determined by the external field $n_{v}^{0}=B/\phi_{0}$ with
$\phi_{0}$ being the quantized flux. Now the ground state of the attractive
$\pi$-flux Hubbard model becomes a QPL with vortex-Fermi-surface, which leads
to quite un-usual physical consequences. People may use the time-of-flight
approach to observe the signature of Fermi liquid of vortices.

\section{Conclusion}

In this paper, based on timely technique, we point out that the realization of
a $\pi$-flux model in a square optical lattice of cold atoms provides an
opportunity to get a new type of quantum fluid. From an attractive $\pi$-flux
Hubbard model, we have an SF/CDW order from spontaneous \textrm{SU(2)}
pseudo-spin rotation symmetry breaking. Due to fairly strong quantum
fluctuations, there may exist a possible quantum phase liquid state, in which
there exists Cooper pairing, but no long range SF phase coherence exists. In
addition, we may even guess that in different SFs with \textrm{SU(2)}
particle-hole symmetry there may exist different types of quantum phase
liquids with different types of SF vortices. This issue will be explored in a
future study.

Finally, we discuss the possible experimental signatures of QPL. The QPL is a
short range quantum SF order. In QPL, the SF correlation decays exponentially
as $\left \langle \hat{\Delta}^{\ast}(x,y)\hat{\Delta}(0)\right \rangle \sim
e^{i\mathbf{q}\cdot \mathbf{x}_{i}}\exp \{-[4\pi v(\frac{1}{g_{c}}-\frac{1}%
{g})]r\}$ with $\mathbf{q}=(\pi,\pi)$. Thus, people may detect the pairing
correlation to observe the QPL. In particular, the fermionic SF vortex in QPL
has different topological properties from the SF vortex in the systems with
pseudo-energy-gap at finite temperature, of which the SF\ vortex always obeys
bosonic statistics. The QPL with finite vortex-density forms a vortex-metal
with Fermi-surface. The Fermi-surface of cold fermions in a 3D optical lattice
has been successfully observed\cite{Fermi surface}. Similarly, one may also
observe these Fermi levels of SF-vortices in this system. In addition, even in
long range SF order (the green region in Fig.5), the\ SF vortex has the same
fermionic zero modes as that in QPL order. People may directly observe the
fermionic zero modes on the SF vortices by time-of-flight imaging in a long
range SF order.

\begin{acknowledgments}
The authors thank\ H. Zhai and R. Q. Wang for helpful discussions. This work
is supported by National Basic Research Program of China (973 Program) under
the grant No. 2011CB921803, 2012CB921704, NSFC Grant No. 11174035.
\end{acknowledgments}

\appendix

\section{Mapping attractive model to repulsive model}

In the Landau gauge, the explicit form of kinetic term of the $\pi$-flux
attractive Hubbard model shown in Eq.(\ref{mod1}) takes the form%
\begin{align}
\hat{H}_{K}^{L}  &  =-\sum_{i,\sigma}t\left(  \hat{c}_{i,\sigma}^{\dagger}%
\hat{c}_{i+e_{x},\sigma}+\hat{c}_{i,\sigma}^{\dagger}\hat{c}_{i-e_{x},\sigma
}\right) \nonumber \\
&  -\sum_{i,\sigma}\left[  \left(  -1\right)  ^{i_{x}}\hat{c}_{i,\sigma
}^{\dagger}\hat{c}_{i+e_{y},\sigma}+\left(  -1\right)  ^{i_{x}}\hat
{c}_{i,\sigma}^{\dagger}\hat{c}_{i-e_{y},\sigma}\right]  .
\end{align}
By the particle-hole transformation $\hat{c}_{i,\uparrow}\rightarrow \tilde
{c}_{i,\uparrow},$\ $\hat{c}_{i,\uparrow}^{\dagger}\rightarrow \tilde
{c}_{i,\uparrow}^{\dagger},\hat{c}_{i,\downarrow}\rightarrow \left(  -1\right)
^{i_{x}+i_{y}}\tilde{c}_{i,\downarrow}^{\dagger},$ $\hat{c}_{i,\downarrow
}^{\dagger}\rightarrow \left(  -1\right)  ^{i_{x}+i_{y}}\tilde{c}%
_{i,\downarrow},$ one may map the attractive Hubbard model to a repulsive
Hubbard model.

The kinetic term becomes
\begin{align}
\hat{H}_{K}^{L}  &  \rightarrow-\sum_{i}t\left(  \tilde{c}_{i,\uparrow
}^{\dagger}\tilde{c}_{i+e_{x},\uparrow}+\tilde{c}_{i,\uparrow}^{\dagger}%
\tilde{c}_{i-e_{x},\uparrow}\right) \nonumber \\
&  -\sum_{i}t\left(  \left(  -1\right)  ^{i_{x}}\tilde{c}_{i,\uparrow
}^{\dagger}\tilde{c}_{i+e_{y},\uparrow}+\left(  -1\right)  ^{i_{x}}\tilde
{c}_{i,\uparrow}^{\dagger}\tilde{c}_{i-e_{y},\uparrow}\right) \nonumber \\
&  -\sum_{i}t\left(  \tilde{c}_{i+e_{x},\downarrow}^{\dagger}\tilde
{c}_{i,\downarrow}+\tilde{c}_{i-e_{x},\downarrow}^{\dagger}\tilde
{c}_{i,\downarrow}\right) \nonumber \\
&  -\sum_{i}t\left(  \left(  -1\right)  ^{i_{x}}\tilde{c}_{i+e_{y},\downarrow
}^{\dagger}\tilde{c}_{i,\downarrow}+\left(  -1\right)  ^{i_{x}}\tilde
{c}_{i-e_{y},\downarrow}^{\dagger}\tilde{c}_{i,\downarrow}\right) \nonumber \\
&  =\tilde{H}_{K}^{L};
\end{align}
The on-site interaction term becomes%
\begin{equation}
\hat{H}_{U}^{L}\rightarrow U\sum \limits_{i}\tilde{n}_{i,\uparrow}\tilde
{n}_{i,\downarrow}-U\sum \limits_{i}\tilde{n}_{i,\uparrow};
\end{equation}
The chemical potential term becomes
\begin{equation}
\hat{H}_{\mu}^{L}\rightarrow-\mu \sum \limits_{i,\sigma}\left(  \tilde
{c}_{i,\uparrow}^{\dagger}\tilde{c}_{i,\uparrow}-\tilde{c}_{i,\downarrow
}^{\dagger}\tilde{c}_{i,\downarrow}\right)  -\mu N;
\end{equation}
The Zeeman field term becomes
\begin{equation}
\hat{H}_{Z}^{L}\rightarrow-h\sum \limits_{i}\left(  \tilde{c}_{i,\uparrow
}^{\dagger}\tilde{c}_{i,\uparrow}+\tilde{c}_{i,\downarrow}^{\dagger}\tilde
{c}_{i,\downarrow}\right)  +hN.
\end{equation}

In summary, the Hamiltonian of the $\pi$-flux attractive Hubbard model under
the particle-hole transformation turns into
\begin{align}
\hat{H}  &  \rightarrow \tilde{H}=\tilde{H}_{K}^{L}+U\sum \limits_{i}\tilde
{n}_{i,\uparrow}\tilde{n}_{i,\downarrow}-\tilde{h}\sum \limits_{i,\sigma
,\sigma^{\prime}}\tilde{c}_{i\sigma}^{\dagger}\sigma_{\sigma \sigma^{\prime}%
}^{z}\tilde{c}_{i\sigma^{\prime}}\nonumber \\
&  -\tilde{\mu}\sum \limits_{i,\sigma}\tilde{n}_{i,\sigma}-\mu N+hN,
\end{align}
where the effective Zeeman field and the effective chemical potential are
given by $\tilde{h}=\mu+\frac{U}{2}$ and $\tilde{\mu}=\frac{U}{2}+h$,
respectively. In the followings, we neglect the constant $-\mu N+hN$, and
arrive at the form of the repulsive Hubbard model as%
\begin{equation}
\tilde{H}=\tilde{H}_{K}^{L}+U\sum \limits_{i}\tilde{n}_{i,\uparrow}\tilde
{n}_{i,\downarrow}-\tilde{\mu}\sum \limits_{i,\sigma}\tilde{n}_{i,\sigma}.
\end{equation}

The order parameters under the particle-hole transformation turns into
\begin{align}
\Delta_{i}  &  =\hat{c}_{i,\uparrow}^{\dagger}\hat{c}_{i,\downarrow}^{\dagger
}\rightarrow \tilde{c}_{i,\uparrow}^{\dagger}\left(  -1\right)  ^{i}\tilde
{c}_{i,\downarrow}\nonumber \\
&  =\left(  -1\right)  ^{i}\hat{\eta}_{i}^{-},\\
\Delta_{i}^{\dagger}  &  =\hat{c}_{i,\downarrow}\hat{c}_{i,\uparrow
}\rightarrow \left(  -1\right)  ^{i}\hat{\eta}_{i}^{+}\\
\rho_{i}  &  =\hat{c}_{i,\uparrow}^{\dagger}\hat{c}_{i,\uparrow}+\hat
{c}_{i,\downarrow}^{\dagger}\hat{c}_{i,\downarrow}\nonumber \\
&  \rightarrow \tilde{c}_{i,\uparrow}^{\dagger}\tilde{c}_{i,\uparrow}-\tilde
{c}_{i,\downarrow}^{\dagger}\tilde{c}_{i,\downarrow}+1\nonumber \\
&  =2\hat{\eta}_{i}^{z}+1,
\end{align}
with $\hat{\eta}^{\pm}=\hat{\eta}_{x}\pm i\hat{\eta}_{y}$, where the
pseudo-spin operators are $\hat{\eta}_{\gamma}=\tilde{c}_{\alpha}^{\dagger
}\sigma_{\alpha,\beta}^{\gamma}\tilde{c}_{\beta}/2$, in which $\sigma^{\gamma
}$ are Pauli matrices with $\gamma=x,y,z$. In conclusion, we have the
relationship between attractive Hubbard model and repulsive Hubbard model as
follows:%
\[%
\begin{tabular}
[c]{cc}\hline \hline
\textbf{Attractive interaction} & \textbf{ Repulsive interaction}\\ \hline
$\left(  -1\right)  ^{i}\Delta_{i}$ & $\hat{\eta}_{i}^{-}$\\
$\left(  -1\right)  ^{i}\Delta_{i}^{\dagger}$ & $\hat{\eta}_{i}^{+}$\\
$\frac{1}{2}\left(  \rho_{i}-1\right)  $ & $\hat{\eta}_{i}^{z}$\\ \hline \hline
\end{tabular}
\]
$\ $

\section{Effective Nonlinear $\sigma$ Model of SF/CDW order}

To study the quantum fluctuations of the SC/CDW order, we get an effective
Lagrangian with spontaneous \textrm{SU(2)} pseudo-spin rotation symmetry
breaking under the particle-hole transformation as
\begin{align}
\mathcal{L}_{\mathrm{eff}}  &  =\sum_{i}\tilde{c}_{i}^{\dagger}\partial_{\tau
}\tilde{c}_{i}-\sum \limits_{\left \langle ij\right \rangle }(t_{i,j}\tilde
{c}_{i}^{\dagger}\tilde{c}_{j}+h.c.)\label{model2}\\
&  -\sum_{i}\left(  -1\right)  ^{i}m_{0}^{HS}\tilde{c}_{i}^{\dagger
}\mathbf{\Delta}_{i}\mathbf{\cdot \sigma}\tilde{c}_{i}-h\sum \limits_{i}%
\tilde{c}_{i}^{\dagger}\tilde{c}_{i}.\nonumber
\end{align}
To describe the quantum fluctuations, we use the Haldane's mapping:%
\begin{align}
\mathbf{\Delta}_{i}  &  =(\operatorname{Re}\Delta_{i},\operatorname{Im}%
\Delta_{i},(\rho_{i}-1)/2)\nonumber \\
&  =\left(  -1\right)  ^{i}\mathbf{n}_{i}\Delta_{0}/2\sqrt{1-\mathbf{L}%
_{i}^{2}}+\mathbf{L}_{i},
\end{align}
where $\mathbf{n}_{i}=(\frac{\operatorname{Re}\Delta_{i}}{\Delta_{0}/2}%
,\frac{\operatorname{Im}\Delta_{i}}{\Delta_{0}/2},\frac{\left(  -1\right)
^{i}(\rho_{i}-1)/2}{\Delta_{0}/2})$ is the \textrm{O(3)} rotor for the SF/CDW
order parameter, which refers to the long wavelength part of $\mathbf{\Delta
}_{i}$ with a restriction $\mathbf{n}_{i}^{2}=1$, and $\mathbf{L}_{i}$ is the
transverse canting field corresponding to the short wavelength part of
$\mathbf{\Delta}_{i}$ with a restriction $\mathbf{L}_{i}\cdot \mathbf{n}_{i}=0$.

We then rotate $\mathbf{\Delta}_{i}$ to the $\mathbf{\hat{z}}$-axis by
performing the following transformation:%
\begin{align}
\Psi_{i}  &  =U_{i}^{\dagger}\tilde{c}_{i},\nonumber \\
U_{i}^{\dagger}\mathbf{n}_{i}.\mathbf{\sigma}U_{i}  &  =\sigma_{z},\nonumber \\
U_{i}^{\dagger}\mathbf{L}_{i}.\mathbf{\sigma}U_{i}  &  =\mathbf{l}%
_{i}.\mathbf{\sigma,}%
\end{align}
where $U_{i}\in \mathrm{SU(2)}/\mathrm{U(1)}.$ One then can derive the
following effective Lagrangian:%
\begin{align}
\mathcal{L}_{\mathrm{eff}}  &  =\sum_{i}\Psi_{i}^{\dagger}\left[
\partial_{\tau}+\left(  U_{i}^{\dagger}\partial_{\tau}U_{i}\right)  -h\right]
\Psi_{i}\nonumber \\
&  -\sum_{\left \langle i,j\right \rangle }\left(  t_{ij}\Psi_{i}^{\dagger
}e^{ia_{ij}}\Psi_{j}+h.c.\right) \nonumber \\
&  -m_{0}^{HS}\sum_{i}\Psi_{i}^{\dagger}\left[  \left(  -1\right)  ^{i}%
\sigma_{z}\sqrt{1-\mathbf{l}_{i}^{2}}+\mathbf{l}_{i}.\mathbf{\sigma}\right]
\Psi_{i},
\end{align}
where $m_{0}^{HS}=U\Delta_{0}/2$, the auxiliary gauge fields $a_{ij}%
=a_{ij,1}\sigma_{x}+a_{ij,2}\sigma_{y},$ and $a_{0}\left(  i\right)
=a_{0,1}\left(  i\right)  \sigma_{x}+a_{0,2}\left(  i\right)  \sigma_{y}$ are
defined as
\begin{align}
e^{ia_{ij}}  &  =U_{i}^{\dagger}U_{j},\nonumber \\
a_{0}\left(  i\right)   &  =U_{i}^{\dagger}\partial_{\tau}U_{i}.
\end{align}

By means of the mean field result $\Delta_{0}/2=(-1)^{i}\left \langle \Psi
_{i}^{\dagger}\sigma_{z}\Psi_{i}\right \rangle $ and the approximations%
\begin{align}
\sqrt{1-\mathbf{l}_{i}^{2}}  &  \simeq1-\frac{\mathbf{l}_{i}^{2}}%
{2},\nonumber \\
e^{ia_{ij}}  &  \simeq1+ia_{ij},
\end{align}
we obtain%
\begin{align}
\mathcal{L}_{\mathrm{eff}}  &  =\sum_{i}\Psi_{i}^{\dagger}\left[
\partial_{\tau}+a_{0}\left(  i\right)  -m_{0}^{HS}\left(  \mathbf{l}%
_{i}.\mathbf{\sigma+}\left(  -1\right)  ^{i}\sigma_{z}\right)  \mathbf{-}%
h\right]  \Psi_{i}\nonumber \\
-  &  \sum_{\left \langle ij\right \rangle }\left[  t_{ij}\Psi_{i}^{\dagger
}\left(  1+ia_{ij}\right)  \Psi_{j}+h.c.\right]  +m_{0}^{HS}\sum_{i}%
\frac{\mathbf{l}_{i}^{2}}{2}.
\end{align}
Performing integration out the fermion field, we then get the effective action%
\begin{align}
\mathcal{S}_{\mathrm{eff}}  &  =\frac{1}{2}\int_{0}^{\beta}d\tau \sum
_{i}\left[  -4\varsigma \left(  a_{0}\left(  i\right)  -m_{0}^{HS}%
\mathbf{l}_{i}.\mathbf{\sigma-}h\right)  ^{2}+4\rho_{\mathrm{phase}}a_{ij}%
^{2}\right] \nonumber \\
&  +\frac{1}{2}\int_{0}^{\beta}d\tau \sum_{i}\frac{2\left(  m_{0}^{HS}\right)
^{2}}{U}\mathbf{l}_{i}^{2}, \label{eq6}%
\end{align}
where $\varsigma$\ and $\rho_{\mathrm{phase}}$ (the phase stiffness) are two
parameters. Next, to learn the properties of the low energy physics, we study
the continuum theory of the effective action in Eq.(\ref{eq6}). In the
continuum limit, we denote the quantities $\mathbf{n}_{i}\rightarrow
\mathbf{n}\left(  x,y\right)  $, $\mathbf{l}_{i}\rightarrow \mathbf{l}\left(
x,y\right)  $, $ia_{ij}=U_{i}^{\dagger}U_{j}-1\rightarrow U^{\dagger}%
\partial_{\mu}U$ ($\mu=x$ or $y$),$\ U_{i}^{\dagger}\partial_{\tau}%
U_{i}\rightarrow U^{\dagger}\partial_{\tau}U$, respectively. From the
relations between $U^{\dagger}\partial_{\mu}U$ and $\partial_{\mu}\mathbf{n}$,
we obtain%
\begin{align}
a_{\tau}^{2}  &  =a_{\tau,1}^{2}+a_{\tau,2}^{2}=-\frac{1}{4}\left(
\partial_{\tau}\mathbf{n}\right)  ^{2},\nonumber \\
a_{\mu}^{2}  &  =a_{\mu,1}^{2}+a_{\mu,2}^{2}=\frac{1}{4}\left(  \partial_{\mu
}\mathbf{n}\right)  ^{2},\mu=x,y,\nonumber \\
\mathbf{a}_{0}.\mathbf{l}  &  \mathbf{=-}\frac{i}{2}\left(  \mathbf{n\times
\partial}_{\tau}\mathbf{n}\right)  .\mathbf{l,}%
\end{align}
where $1$, $2$ denote the two spin flavors. We then integrate out the
transverse canting field $\mathbf{l}$ and obtain the effective action as
follows:%
\begin{equation}
\mathcal{S}_{\mathrm{eff}}=\frac{\rho_{\mathrm{phase}}}{2}\int_{0}^{\beta
}d\tau \int d^{2}r\left[  \left(  \bigtriangledown \mathbf{n}\right)  ^{2}%
+\frac{1}{v^{2}}\left(  \mathbf{\partial}_{\tau}\mathbf{n}\right)
^{2}\right]  +S_{B}\left[  \mathbf{n}\right]  ,
\end{equation}
where $v=\sqrt{\rho_{\mathrm{phase}}\left(  1/\varsigma-2U\right)  }$ and
$S_{B}\left[  \mathbf{n}\right]  $ are other terms which are irrelevant to the
second order term about vector $\mathbf{n}$\textbf{.}

To give the coefficients $\varsigma$ and $\rho_{s}$, we choose $U_{i}$ in
\textrm{CP(1)} representation to be
\begin{equation}
U_{i}=\left(
\begin{array}
[c]{cc}%
z_{i\uparrow}^{\ast} & z_{i\downarrow}^{\ast}\\
-z_{i\downarrow} & z_{i\uparrow}%
\end{array}
\right)  ,
\end{equation}
where $\mathbf{z}_{i}=\left(  z_{i\uparrow},z_{i\downarrow}\right)  ^{T},$
$\mathbf{\bar{z}}_{i}\mathbf{z}_{i}=1$, and $\mathbf{n}_{i}=\mathbf{\bar{z}%
}_{i}\mathbf{\sigma z}_{i}$ \cite{wen}. The quantum fluctuations around
$\mathbf{n}_{i}=\mathbf{\hat{z}}_{i}$ is
\begin{align}
\mathbf{n}_{i}  &  =\mathbf{\hat{z}}_{i}+\operatorname{Re}\left(  \phi
_{i}\right)  \mathbf{\hat{x}}_{i}+\operatorname{Im}\left(  \phi_{i}\right)
\mathbf{\hat{y}}_{i},\nonumber \\
\mathbf{\hat{z}}_{i}  &  =\left(
\begin{array}
[c]{c}%
1-\frac{\left \vert \phi_{i}\right \vert ^{2}}{8}\\
\frac{\phi_{i}}{2}%
\end{array}
\right)  +O\left(  \phi_{i}^{3}\right)  .
\end{align}
Then the quantities $U_{i}^{\dagger}U_{j}$ and $U_{i}^{\dagger}\partial_{\tau
}U_{i}$ can be expanded in the power of $\phi_{i}-\phi_{j}$ and $\partial
_{\tau}\phi_{i}$, i.e.,%
\begin{align}
U_{i}^{\dagger}U_{j}  &  =e^{-\frac{i}{2}\left(  \phi_{i}-\phi_{j}\right)
}\sigma_{y},\nonumber \\
U_{i}^{\dagger}\partial \tau U_{i}  &  =\left(
\begin{array}
[c]{cc}%
0 & \frac{1}{2}\partial_{\tau}\phi_{i}\\
-\frac{1}{2}\partial_{\tau}\phi_{i} & 0
\end{array}
\right)  .
\end{align}
The gauge field $a_{ij}$ and $a_{0}(i)$ are therefore given by%
\begin{align}
a_{ij}  &  =-\frac{i}{2}\left(  \phi_{i}-\phi_{j}\right)  ,\nonumber \\
a_{0}\left(  i\right)   &  =-\frac{i}{2}\partial_{\tau}\phi_{i}.
\end{align}
Assuming that $a_{ij}$ and $a_{0}(i)$ are constant in space and denoting
$\partial_{i}\phi_{i}=\mathbf{a}$, and $\partial_{\tau}\phi_{i}=iB_{y}$, we
get%
\begin{align}
a_{ij}  &  =-\frac{1}{2}\mathbf{a.}\left(  \mathbf{i}-\mathbf{j}\right)
\sigma_{y},\nonumber \\
a_{0}(i)  &  =-\frac{1}{2}B_{y}\sigma_{y}.
\end{align}
The energy of Hamiltonian of Eq.(\ref{eq6}) becomes%
\begin{equation}
E\left(  B_{y},\mathbf{a}\right)  =-\frac{1}{2}\varsigma B_{y}^{2}+\frac{1}%
{2}\rho_{\mathrm{phase}}\mathbf{a}^{2}.
\end{equation}
Then one may obtain $\varsigma$ and $\rho_{\mathrm{phase}}$ by the partial
derivative of the energy%
\begin{align}
\varsigma &  =-\frac{1}{N}\frac{\partial^{2}E_{0}\left(  B_{y}\right)
}{\partial B_{y}^{2}}\left \vert _{B_{y}=0},\right. \nonumber \\
\rho_{\mathrm{phase}}  &  =\frac{1}{N}\frac{\partial^{2}E_{0}\left(
\mathbf{a}\right)  }{\partial \mathbf{a}^{2}}\left \vert _{\mathbf{a}%
=0},\right.
\end{align}
where $N=2N_{s}$. Here $E_{0}(B_{y})$ and $E_{0}$ $(\mathbf{a})$ are the
energy spectra of the lower Hubbard band%
\begin{align}
E_{0}(B_{y})  &  =\sum_{\mathbf{k}}\left(  E_{+,\mathbf{k}}^{\varsigma
}+E_{-,\mathbf{k}}^{\varsigma}\right)  ,\nonumber \\
E_{0}(\mathbf{a})  &  =\sum_{\mathbf{k}}\left(  E_{+,\mathbf{k}}%
^{\rho_{\mathrm{phase}}}+E_{-,\mathbf{k}}^{\rho_{\mathrm{phase}}}\right)  ,
\end{align}
where $E_{+,\mathbf{k}}^{\varsigma},$ $E_{-,\mathbf{k}}^{\varsigma}$ and
$E_{+,\mathbf{k}}^{\rho_{\mathrm{phase}}}$, $E_{-,\mathbf{k}}^{\rho
_{\mathrm{phase}}}$ are the energy spectra of the following Hamiltonian
$H^{\varsigma}$ and $H^{\rho_{\mathrm{phase}}}$ given by%
\begin{align}
H^{\varsigma}  &  =-\sum_{\left \langle i,j\right \rangle }\left(  t_{ij}%
\Psi_{i}^{\dagger}\Psi_{j}+h.c.\right)  -m_{0}^{HS}\sum_{i}\Psi_{i}^{\dagger
}\left(  -1\right)  ^{i}\sigma_{z}\Psi_{i}\nonumber \\
&  -\sum_{i}\frac{B_{y}}{2}\Psi_{i}^{\dagger}\sigma_{y}\Psi_{i}-h\sum_{i}%
\Psi_{i}^{\dagger}\Psi_{i},
\end{align}%
\begin{align}
H^{\rho_{\mathrm{phase}}}  &  =-\sum_{\left \langle i,j\right \rangle }\left(
t_{ij}\Psi_{i}^{\dagger}e^{ia_{ij}}\Psi_{j}+h.c.\right)  -h\sum_{i}\Psi
_{i}^{\dagger}\Psi_{i}\nonumber \\
&  -m_{0}^{HS}\sum_{i}\Psi_{i}^{\dagger}\left(  -1\right)  ^{i}\sigma_{z}%
\Psi_{i},
\end{align}
where $a_{ij}=\frac{1}{2}\left(  \mathbf{i}-\mathbf{j}\right)  \cdot
\mathbf{\sigma}$.

By the Fourier transformation for $H^{\varsigma}$, we get the spectra of
$H^{\varsigma}$:%
\begin{equation}
E_{\pm,\mathbf{k}}^{\varsigma}=-\sqrt{\left(  \left \vert \xi_{k}\right \vert
\pm \frac{B_{y}}{2}\right)  ^{2}+\left(  m_{0}^{HS}\right)  ^{2}}.
\end{equation}
Making use of $\varsigma=-\frac{1}{N}\frac{\partial^{2}E_{0}\left(
B_{y}\right)  }{\partial B_{y}^{2}}\mid_{B_{y}=0},$ and $E_{0}\left(
B_{y}\right)  =\sum_{\mathbf{k}}\left(  E_{+,\mathbf{k}}^{\varsigma
}+E_{-,\mathbf{k}}^{\varsigma}\right)  ,$ we can obtain%
\begin{equation}
\varsigma=\frac{1}{4N_{s}}\sum_{E_{\mathbf{k}}>-h}\frac{\left(  m_{0}%
^{HS}\right)  ^{2}}{\left(  \left \vert \xi_{k}\right \vert ^{2}+\Delta
^{2}\right)  ^{\frac{3}{2}}}.
\end{equation}
Similarly, we can get energy spectra of $H^{\rho_{\mathrm{phase}}}$:
\begin{equation}
E_{\pm,\mathbf{k}}^{\rho_{\mathrm{phase}}}=-\sqrt{P_{x}^{2}+P_{y}^{2}%
+Q_{x}^{2}+Q_{y}^{2}\pm2f},
\end{equation}
where
\begin{align}
f  &  =\sqrt{\Delta^{2}\left(  Q_{x}^{2}+Q_{y}^{2}\right)  +\left(  P_{x}%
Q_{x}+P_{y}Q_{y}\right)  ^{2}},\nonumber \\
P_{x}  &  =2t\cos \left(  \frac{a_{x}}{2}\right)  \cos k_{x},\nonumber \\
P_{y}  &  =2t\cos \left(  \frac{a_{y}}{2}\right)  \cos k_{y},\nonumber \\
Q_{x}  &  =2t\sin \left(  \frac{a_{x}}{2}\right)  \sin k_{x},\nonumber \\
P_{y}  &  =2t\sin \left(  \frac{a_{y}}{2}\right)  \sin k_{y}.
\end{align}

Finally, we derive the effective \textrm{O(3)} nonlinear $\sigma$-model
(NL$\sigma$M):
\begin{equation}
\mathcal{L}_{\mathrm{SF/CDW}}=\frac{1}{2g}[\left(  \partial_{\tau}%
\mathbf{n}\right)  ^{2}+\frac{1}{v^{2}}\left(  \mathbf{\bigtriangledown
n}\right)  ^{2}].
\end{equation}
The coupling constant $g$ and the collective mode's velocity $v$ are defined
as
\begin{align}
g  &  =\frac{v}{\rho_{\mathrm{phase}}},\nonumber \\
v^{2}  &  =\rho_{\mathrm{phase}}[(\frac{1}{4N_{s}}\sum_{E_{\mathbf{k}}%
>-h}\frac{(m_{0}^{HS})^{2}}{E_{\mathbf{k}}^{\frac{3}{2}}})^{-1}-2U]
\end{align}
where $m_{0}^{HS}=U\Delta_{0}/2$, and the phase stiffness $\rho
_{\mathrm{phase}}$ of the SF order is shown in Eq.(\ref{rou}) in the main text.


\begin{thebibliography}{99}                                                                                               %


\bibitem {Bloch's review}I. Bloch, J. Dalibard and W. Zwerger, Rev. Mod. Phys.
\textbf{80}, 885 (2008).

\bibitem {review}S. Giorgini, L. P. Pitaevskii and S. Stringari, Rev. Mod.
Phys. \textbf{80}, 1215 (2008).

\bibitem {fesh}T. K\"{o}hler, K. G\'{o}ral, and P. S. Julienne, Rev. Mod.
Phys. \textbf{78}, 1311-1361 (2006).

\bibitem {fesh1}C. Chin, R. Grimm, P. Julienne and E. Tiesinga, Rev. Mod.
Phys. \textbf{82}, 1225-1286 (2010).

\bibitem {ai}M. Aidelsburger, M. Atala, S. Nascimb\`{e}ne, S. Trotzky, Y.-A.
Chen, and I. Bloch, Phys. Rev. Lett. \textbf{107,} 255301 (2011).

\bibitem {ai1}M. Aidelsburger, M. Atala, M. Lohse, J. T. Barreiro, B. Paredes,
I. Bloch, Phys. Rev. Lett. \textbf{111}, 185301 (2013).

\bibitem {fl}H. Miyake, G. A. Siviloglou, C. J. Kennedy, W. Cody Burton, W.
Ketterle, Phys. Rev. Lett. \textbf{111}, 185302 (2013).

\bibitem {zhai}Hui Zhai, R.O. Umucahlar, and M.\"{O}. Oktel, Phys. Rev. Lett.
\textbf{104}, 145301 (2010).

\bibitem {mic}C. N. Yang and S. C. Zhang, Mod. Phys. Lett. \textbf{B 4,} 759 (1990).

\bibitem {Iskin}M. Iskin and C. A. R. S\'{a} de Melo, Phys. Rev. B
\textbf{72}, 024512 (2005).

\bibitem {Haldane}F. D. M. Haldane, Phys. Rev. Lett. \textbf{61}, 2015 (1988).

\bibitem {Dupuis}N. Dupuis, Phys. Rev. \textbf{B 70}, 134502 (2004).

\bibitem {cha}S. Chakravarty, Bertrand I. Halperin and David R. Nelson, Phys.
Rev. B \textbf{39}, 2344 (1989).

\bibitem {sech}S. Sachdev, \emph{Quantum Phase Transitions}, (Cambridge
University Press, 1999).

\bibitem {exp}In current experiment, for example, according to parameters from
Ref.\cite{ai}, the order of $k_{B}T=0.02t$ is about \textrm{nK. }This
low-temperature-condition is still a challenge for the fermionic system.

\bibitem {kou1}G. Y. Sun and S. P. Kou, EPL, \textbf{87,} 67002 (2009).

\bibitem {pi}C. C. Chang and R. T. Scalettar, Phys. Rev. Lett. \textbf{109},
026404 (2012).

\bibitem {kou3}S. P. Kou, Phys. Rev. \textbf{B 78}, 233104 (2008).

\bibitem {comment}The QPL is a gapped state. Thus, the properties are robust
to the perturbations. The small variation of the (uniform) $\pi$-flux will
cause additional fermionic SF-vortices. If there exists a small extra magnetic
field away from the magnetic field in $\pi$-flux case, the density of the
fermionic SF vortices becomes finite. When the extra magnetic field is far
away from the magnetic field in $\pi$-flux case, the femionic system has
different magnetic translation symmetry and different mean field ansatz. So,
in this paper, we only consider the case of magnetic field away from that in
$\pi$-flux case slightly.

\bibitem {Fermi surface}M. K\"{o}hl, H. Moritz, T. St\"{o}ferle, K. G\"{u}nter
and T. Esslinger, Phys. Rev. Lett. \textbf{94}, 080403 (2005).

\bibitem {wen}X. G. Wen, Quantum Field Theory of Many-Body Systems (Oxford
Univ. Press, Oxford, 2004).
\end{thebibliography}
\end{document}